\documentclass[12pt A4paper]{article}
\usepackage{latexsym}
\usepackage{graphicx}
\usepackage{amsmath}
\usepackage{amssymb}
\graphicspath{{./}{./figs/}}

\begin{document}

\title{Stochastic Synchronization of Genetic Oscillator Networks\thanks{BMC Systems Biology, Vol.1, article no.6, 2007. doi:10.1186/1752-0509-1-6}}
\author{Chunguang Li$^{1,2,3}$, Luonan Chen$^{2,3,4,5}$, Kazuyuki Aihara$^{2,3}$}
\date{\small
$^1$Centre for Nonlinear and Complex Systems, University of
Electronic Science and Technology of China, Chengdu
610054, P. R. China \\
$^2$ERATO Aihara Complexity Modelling Project, Room M204, Komaba
Open Laboratory, University of Tokyo,
4-6-1 Komaba, Meguro-ku, Tokyo 153-8505, Japan\\
$^3$Institute of Industrial Science, University of Tokyo, Tokyo 153-8505, Japan\\
$^4$Institute of Systems Biology, Shanghai University, P. R.
China\\
$^5$Department of Electrical Engineering and Electronics,
Osaka Sanyo University, Osaka, Japan}

\maketitle

\begin{abstract}
{\bf Background}: The study of synchronization among genetic
oscillators is essential for the understanding of the rhythmic
phenomena of living organisms at both molecular and cellular
levels. Genetic networks are intrinsically noisy due to natural
random intra- and inter-cellular fluctuations. Therefore, it is
important to study the effects of noise perturbation on the
synchronous dynamics of genetic oscillators. From the synthetic
biology viewpoint, it is also important to implement biological
systems that minimizing the negative influence of the
perturbations.

{\bf Results}: In this paper, based on systems biology approach,
we provide a general theoretical result on the synchronization of
genetic oscillators with stochastic perturbations. By exploiting
the specific properties of many genetic oscillator models, we
provide an easy-verified sufficient condition for the stochastic
synchronization of coupled genetic oscillators, based on the Lur'e
system approach in control theory. A design principle for
minimizing the influence of noise is also presented. To
demonstrate the effectiveness of our theoretical results, a
population of coupled repressillators is adopted as a numerical
example.

{\bf Conclusion}: In summary, we present an efficient theoretical
method for analyzing the synchronization of genetic oscillator
networks, which is helpful for understanding and testing the
synchronization phenomena in biological organisms. Besides, the
results are actually applicable to general oscillator networks.

\end{abstract}
\section*{Background}
Elucidating the collective dynamics of coupled genetic oscillators
not only is important for the understanding of the rhythmic
phenomena of living organisms, but also has many potential
applications in bioengineering areas. So far, many researchers
have studied the synchronization in genetic networks from the
aspects of experiment, numerical simulation and theoretical
analysis. For instance, in \cite{1}, the authors experimentally
investigated the synchronization of cellular clock in the
suprachiasmatic nucleus (SCN); in \cite{2a, 2b, 2c}, the
synchronization are studied in biological networks of identical
genetic oscillators; and in \cite{3,4,5}, the synchronization for
coupled nonidentical genetic oscillators is investigated. Gene
regulation is an intrinsically noisy process, which is subject to
intracellular and extracellular noise perturbations and
environment fluctuations \cite{6a, 6b, 6c, 6d, 6e, plos}. Such
cellular noises will undoubtedly affect the dynamics of the
networks both quantitatively and qualitatively. In \cite{7}, the
authors numerically studied the cooperative behaviors of a
multicell system with noise perturbations. But to our knowledge,
the synchronization properties of stochastic genetic networks have
not yet been theoretically studied.

This paper aims to provide a theoretical result for the
synchronization of coupled genetic oscillators with noise
perturbations, based on control theory approach. We first provide
a general theoretical result for the stochastic synchronization of
coupled oscillators. After that, by taking the specific structure
of many model genetic oscillators into account, we present a
sufficient condition for the stochastic synchronization in terms
of linear matrix inequalities (LMIs) \cite{13}, which are very
easy to be verified numerically. To our knowledge, the
synchronization of complex oscillator networks with noise
perturbations, even not in the biological context, has not yet
been fully studied. Recently, it was found that many biological
networks are complex networks with small-world and scale-free
properties \cite{14a, 14b}. Our method is also applicable to
genetic oscillator networks with complex topology, directed and
weighted couplings. To demonstrate the effectiveness of the
theoretical results, we present a simulation example of coupled
repressilators. Throughout this paper, matrix $U\in R^{N\times N}$
is defined as an irreducible matrices with zero row sums, whose
off-diagonal elements are all non-positive, and the other
notations are defined in the Appendix A.

\section*{Results}
\subsection*{Theoretical Results}
Since we know very little about how the cellular noises act on the
genetic networks, a simple way to incorporate random effects is to
assume that certain noises randomly perturb the genetic networks
in an additive manner. We consider the following networks of $N$
coupled genetic oscillators with random noise perturbations
\begin{equation*}
\frac{dx_i(t)}{dt}=F(x_i(t))+\sum_{j=1}^NG_{ij}Dx_j(t)
+v_i(t)n_i(t),\,i=1,\cdots,N
\end{equation*}
where $F(.)$ defines the dynamics of individual oscillators,
$v_i(t)\in R^{n\times 1}$ is called the noise intensity vector,
belongs to $L_2[0,\infty)$.  As we will see in the following
analysis, the results hold no matter what $v_i(t)$ is and no
matter where it is introduced, so we do not explicitly express the
form of $v_i(t)$ here. $v_i(t)$ can also be a function of the
variables (if so, some minor modifications are needed in the
following). $n_i(t)$ is a scalar zero mean Gaussian white noise
process. Recall that the time derivative of a Wiener process is a
white noise process \cite{Arnold}, hence we can define
$dw_i(t)=n_i(t)dt$, where $w_i(t)$ is a scalar Wiener process.
Thus, the above equation can be rewritten as the following
stochastic differential equation form:
\begin{equation}
dx_i(t)=\left[F(x_i(t))+\sum_{j=1}^NG_{ij}Dx_j(t)\right]dt
+v_i(t)dw_i(t),\,i=1,\cdots,N.
\end{equation}
The work can be easily extended to the case that $v_i(t)\in
R^{n\times l_i}$ and $n_i(t)=[n_{i1}(t),\cdots,n_{il_i}(t)]^T$ be
an $l_i$-dimensional mutually independent zero mean Gaussian white
noise process. $D\in R^{n\times n}$ defines the coupling between
two genetic oscillators. $G=(G_{ij})_{N\times N}$ is the coupling
matrix of the network. If there is a link from oscillator $j$ to
oscillator $i$ $(j\neq i)$, then $G_{ij}$ equals to a positive
constant denoting the coupling strength of this link; otherwise,
$G_{ij}=0$; $G_{ii}=-\sum_{j=1, j\neq i}^N G_{ij}$. Matrix $G$
defines the coupling topology, direction, and the coupling
strength of the network.

For network (1), a natural attempt is to study the mean-square
asymptotic synchronization. But  existing experimental results
show that usually the genetic oscillators can not achieve
mean-square synchronization (see, e.g. experimental results in [1]
and Appendix B for a theoretical discussion). Analogue to the
stochastic stability with disturbance attenuation (see, e.g.
\cite{Xu}), we give a less restrictive (but more realistic)
definition of the stochastic synchronization as follows:

\emph{Definition 1}: For a given scalar $\gamma>0$, the network
(1) is said to be stochastically synchronous (under the
combination matrix $U$) with disturbance attenuation $\gamma$ if
the network without disturbance $(v_i=0, \forall i)$ is
asymptotically synchronous, and under the same initial conditions
for all oscillators,
\begin{equation}
\sum_{i<j}(-U_{ij})\|x_i(t)-x_j(t)\|_{E_2}^2<\gamma
\sum_i\|v_i(t)\|_2^2,
\end{equation}
for all nonzero $v_i(t)$, where $\|\cdot\|_{E_2}^2=\textbf{E}
\left(\int_0^\infty |\cdot|^2 dt\right)$. Here, by introducing the
combination matrix $U$, we can flexibly select the form of the
matrix to obtain different error combinations.

By using the techniques described in the Appendix C, we know that
if there exist matrices $P>0, T\in R^{n\times n}$ and $U$, and a
scalar $\rho>0$, such that the following conditions are satisfied,
\begin{equation}
\begin{array}{c}
S_2\equiv 2(y_1-y_2)^TP[F(y_1)-F(y_2)-T(y_1-y_2)]
+\frac{\rho}{\gamma}(y_1-y_2)^T (y_1-y_2)<0,\,  \forall y_1,y_2 \in R^n\, (y_1\neq y_2),\\
(U\otimes P)(G\otimes D+I\otimes T)+(G\otimes D+I\otimes
T)^T(U\otimes P)\leq 0,\\
U\otimes P\leq \rho I,
\end{array}
\end{equation}
then, the network (1) will achieve stochastic synchronization with
disturbance attenuation $\gamma$.

The above condition (3) is a general result for the stochastic
synchronization of coupled oscillators. But we do not have general
efficient method for verifying the first inequality in (3) for
arbitrary $F$ due to its nonlinearity. Next we consider a special
structure of genetic oscillators to obtain an easy-verified
result.

Genetic oscillators are biochemically dynamical networks, which
can usually be modelled as nonlinear dynamical systems.
Mathematically many genetic oscillators can be expressed in the
form of multiple additive terms, and the terms are monotonic
functions of each variable, which particularly, are of linear,
Michaelis-Menten and Hill forms. In our previous papers \cite{5,
TCAS}, we have taken such special structure properties of gene
networks into account, and have shown that these genetic
oscillators can be transformed into Lur'e form and can be further
analyzed by using Lur'e system method in control theory \cite{12}.
In this paper, we also consider such special structure. To make
our paper more understandable and self-contained, we will first
introduce the approach briefly, and after that we will analyze the
stochastic genetic oscillator networks theoretically. We consider
the following general form of genetic oscillator:
\begin{equation}
\dot{y}(t)=Ay(t)+B_1 f_1(y(t))+B_2 f_2(y(t)),
\end{equation}
where $y(t)\in R^n$ represents the concentrations of proteins,
RNAs and chemical complexes, $A$, $B_1$ and $B_2$ are matrices in
$R^{n\times n}$, $f_1(y(t))=
[f_{11}(y_1(t)),\cdots,f_{1n}(y_n(t))]^T$ with $f_{1j}(y_j(t))$ as
a monotonic increasing function of the form $
f_{1j}(y_j(t))=(y_j(t)/\beta_{1j})^{H_{1j}}/[1+(y_j(t)/\beta_{1j})^{H_{1j}}]$,
and $f_2(y(t))= [f_{21}(y_1(t)),\cdots,f_{2n}(y_n(t))]^T$ with
$f_{2j}(y_j(t))$ as a monotonic decreasing function of the form
$f_{2j}(y_j(t))=1/[1+(y_j(t)/\beta_{2j})^{H_{2j}}]$, where
$H_{1j}$ and $H_{2j}$ are the Hill coefficients. Genetic
oscillators of the form (4) is by no mean peculiar. Many
well-known genetic system models can be represented in this form,
such as the Goodwin model \cite{8}, the repressilator \cite{9},
the toggle switch \cite{10}, and the circadian oscillators
\cite{11}. Undoubtedly, this work can be easily generalized to the
case of $\dot{y}(t)=Ay(t)+\sum_{j=1}^l B_j f_j(y(t))$, where there
are more than two nonlinear terms in each equation of the genetic
oscillator. From the synthetic biology viewpoint, genetic
oscillators with only linear, Michaelis-Menten and Hill terms can
also be implemented experimentally.

To avoid confusion, we let the $j$th column of $B_{1,2}$ be zeros
if $f_{1j,2j}\equiv 0$. Since $
f_{2j}(y_j(t))=\frac{1}{1+(y_j(t)/\beta_{2j})^{H_{2j}}}
=1-\frac{(y_j(t)/\beta_{2j})^{H_{2j}}}{1+(y_j(t)/\beta_{2j})^{H_{2j}}}
\equiv 1-g_{j}(y_j(t))$, and letting $f(.)=f_1(.)$, we can rewrite
(4) as follows:
\begin{equation}
\dot{y}(t)=Ay(t)+B_1 f(y(t))-B_2 g(y(t))+B_2.
\end{equation}
Obviously, $f_i$ and $g_i$ satisfy the sector conditions: $0\leq
\frac{f_i(a)-f_i(b)}{a-b} \leq k_{1i}, 0\leq
\frac{g_i(a)-g_i(b)}{a-b} \leq k_{2i}$, or equivalently,
\begin{equation}
\begin{array}{c}
(f_i(a)-f_i(b))[(f_i(a)-f_i(b))-k_{1i}(a-b)] \leq 0,\\
(g_i(a)-g_i(b))[(g_i(a)-g_i(b))-k_{2i}(a-b)] \leq 0,\\
\forall a, b\in R\, (a\neq b); i=1,\cdots,n,
\end{array}
\end{equation}

Recall that a Lur'e system is a linear dynamic system, feedback
interconnected to a static nonlinearity that satisfies a sector
condition \cite{12}. Hence, the genetic oscillator (5) can be seen
as a Lur'e system, which can be investigated by using the fruitful
Lur'e system approach in control theory.

By substituting the individual genetic oscillator dynamics (5) for
$F$ in the network (1), we obtain the following network of $N$
coupled genetic oscillators:
\begin{equation}
\begin{array}{rl}
dx_i(t)=&[Ax_i(t)+B_1 f(x_i(t))-B_2
g(x_i(t))+B_2+\sum_{j=1}^NG_{ij}Dx_j(t)]dt+v_i(t)dw_i(t),\\
&i=1,\cdots,N.
\end{array}
\end{equation}

For this network, we have the following result:

\emph{Proposition 1}: If there are matrices  $P>0$,
$\Lambda_1=\mbox{diag}(\lambda_{11},\cdots,\lambda_{1n})>0$,
$\Lambda_2=\mbox{diag}(\lambda_{21},\cdots,\lambda_{2n})>0$, $Q\in
R^{n\times n}$, $U$ as defined above, and a positive real constant
$\rho$ such that the following matrix inequalities hold
\begin{equation}
\begin{array}{l}
\left[\begin{array}{ccc}(1,1) &PB_1+K_1\Lambda_1
&-PB_2+K_2\Lambda_2\\B_1^TP+K_1\Lambda_1&-2\Lambda_1&0 \\
-B_2^TP+K_2\Lambda_2&0&-2\Lambda_2\end{array}\right]<0,\\
(UG\otimes PD+U\otimes Q)^T+(UG\otimes PD+U\otimes Q)\leq 0,\\
U\otimes P\leq \rho I,
\end{array}
\end{equation}
where $(1,1)=PA+A^TP-Q-Q^T+\frac{\rho}{\gamma}I$,
$K_1=\mbox{diag}(k_{11},\cdots,k_{1n}),
K_2=\mbox{diag}(k_{21},\cdots,k_{2n})$. Then the network (7) is
stochastic synchronization with disturbance attenuation $\gamma$.

Proposition 1 can be proved by replacing $F$ in $S_2$ of (3) by
the dynamics of (5), and using the sector conditions (6). The
details are given in Appendix D. If we choose $U$ beforehand, the
matrix inequalities in (8) are all LMIs, which are very easy to be
verified numerically \cite{13}. For some special $G$ and $D$, we
can further simplify the verification process \cite{Wu, 5}.

\subsection*{An Example}

To demonstrate the effectiveness of our theoretical results, we
consider a population of $N$ coupled biological clocks, and the
individual genetic oscillator is the repressilator \cite{9}. The
repressilator is a network of three genes, the products of which
inhibit the transcription of each other in a cyclic way (10).
Specifically, the gene {\it lacI} expresses protein LacI, which
inhibits transcription of the gene {\it tetR}. The protein product
TetR, inhibits transcription of the gene {\it cI}, the protein
product CI of which in turn inhibits expression of {\it lacI},
thus forming a negative feedback cycle.

The quorum-sensing system is used for the coupling purpose, which
was described in \cite{3}. The system achieves cell-to-cell
communication through a mechanism that makes use of two proteins,
the first one of which (LuxI), under the control of the
repressilator protein LacI, synthesizes a small molecule known as
an autoinducer (AI), which can diffuse freely through the cell
membrane. When a second protein (LuxR) binds to this molecule, the
resulting complex activates the transcription {\it lacI}, as shown
in Fig. 1. The noise perturbations in the model can arise both
intracellularly, due to the intrinsically noisy property of the
gene regulation process, and extracellularly, due to environment
fluctuations.
\begin{figure}[htb]
\centering
\includegraphics[width=15cm]{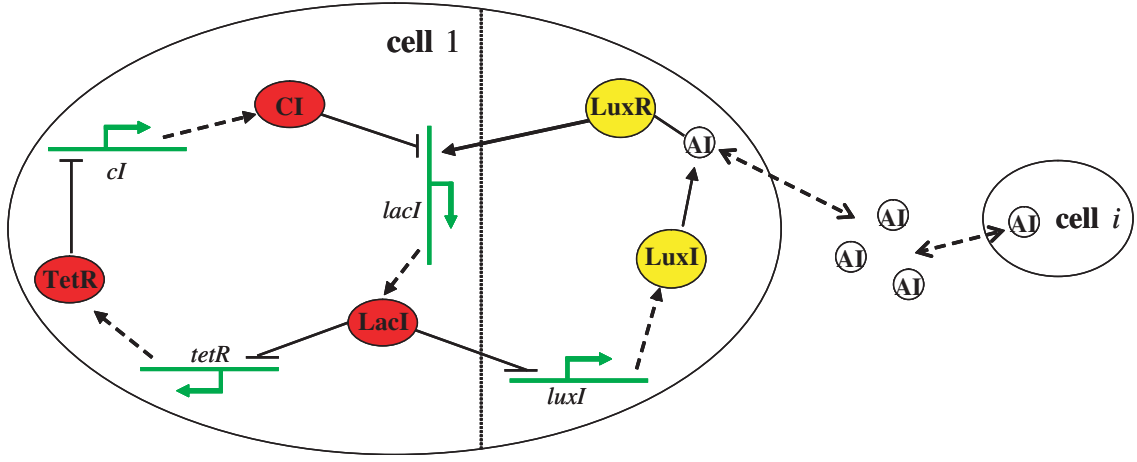}
\caption{Schematic representation of the coupled repressilator
network. In the left big circle, detailed regulation and coupling
mechanism are presented. The repressilator module is located at
the left of the vertical dotted line, and the coupling module
appears at the right.}
\end{figure}

To model the system, we use $a_i, b_i, c_i$ and $A_i,B_i,C_i$ to
represent the dimensionless concentrations of the genes {\it tetR,
cI, lacI} and their product proteins TetR, CI, LacI, respectively.
As in \cite{3}, assuming equal lifetimes of the TetR and LuxI
proteins, their dynamics are identical, and hence we can use the
same variable to describe both protein concentrations. The
concentration of AI inside the $i$th cell is denoted by $S_i$.
Consequently, the mRNA and protein dynamics in the $i$th cell can
be described by \cite{3}:
\begin{equation}
\begin{array}{l}
\frac{da_i}{dt}=-d_1a_i+\frac{\alpha}{\mu+C_i^m},\\
\frac{db_i}{dt}=-d_2b_i+\frac{\alpha}{\mu+A_i^m},\\
\frac{dc_i}{dt}=-d_3c_i+\frac{\alpha}{\mu+B_i^m}+\frac{kS_i}{\mu_s+S_i},\\
\frac{dA_i}{dt}=-d_4A_i+\beta_1 a_i,\\
\frac{dB_i}{dt}=-d_5B_i+\beta_2 b_i,\\
\frac{dC_i}{dt}=-d_6C_i+\beta_3 c_i,\\
\frac{dS_i}{dt}=-k_{s0}S_i+k_{s1} A_i+\eta(S_e-S_i),
\end{array}
\end{equation}
where $m$ is the Hill coefficient, $S_e$ denotes the extracellular
AI concentration, and the meaning of the other parameters are
standard in genetic network models. We assume that the release of
the AI is fast with respect to the timescale of the oscillators
and becomes approximately homogeneous to establish an average AI
level outside the cells. In the quasi-steady-state approximation,
the extracellular AI concentration can be approximated by \cite{3}
\begin{equation*}
S_e=Q_0\bar{S}=\frac{Q_0}{N}\sum_{j=1}^NS_j,
\end{equation*}
where $0<Q_0<1$ is a constant. Thus the dynamics of $S_i$ can be
rewritten as
\begin{equation}
\frac{dS_i}{dt}=-[k_{s0}+(1-Q_0)\eta]S_i+k_{s1} A_i+\frac{\eta
Q_0}{N}\sum_{j=1}^N(S_j-S_i)
\end{equation}

Clearly, the individual model in (9) is of the form (5), in which
$f=[0,0,0,0,0,0,S_i/(\mu_s+S_i)]^T$, $B_1$ is a $7 \times 7$
matrix with all zero entries except for $B_1(3,7)=k$, $g=[0,0,0,
A_i^m/(\mu+A_i^m),B_i^m/(\mu+B_i^m),C_i^m/(\mu+C_i^m),0]^T$, $B_2$
is a $7 \times 7$ matrix with all zero entries except for
$B_2(1,6)=B_2(2,4)=B_2(3,5)=\alpha/\mu$, and all the other terms
are in linear form. Obviously, the coupling term can also be
written into the form defined previously.

The purpose of this example is to demonstrate the effectiveness
and correctness of the theoretical result, instead of mimicking
the real biological clock system. We consider a small size of
network with $N=10$ coupled oscillators. The parameters are set as
$m=4, \alpha=1.8, d_{1}=d_{2}=d_{3}=0.4, \mu=1.3, k=5, \mu_s=5$,
$d_{4}=d_{5}=d_{6}=0.5, \beta_1=\beta_2=\beta_3=0.2, k_{s0}=0.016,
k_{s1}=0.018$, $Q_0=0.8$ and $\eta=0.4$. Since Proposition 1 holds
no matter what $v_i(t)$ is and no matter where it is introduced,
and the verification of Proposition 1 is independent of noise
intensity $v_i$, for simplicity, we set $v_i=0.015$ as a scalar
for all $i$, and the noise term $v_in_i(t)$ is added to the first
equation in (9), where $n_i(t)$ is a scalar Gaussian white noise
process. According to Proposition 1 (by letting $U=-G$, and using
MATLAB LMI Toolbox), we know that the above all-to-all coupled
network can achieve stochastic synchronization with disturbance
attenuation $\gamma=6$. Although $\gamma$ is a large value, it is
easy to show from (2) that the time average of
\textbf{E}($\sum_i\sum_j {|x_i(t)-x_j(t)|}^2$) is still rather
small because $v_i^2$ is very small. We omit the computational
details here. In Fig. 1 (a)\& (b), when starting from the same
initial values, we plot the time evolution of the mRNA
concentrations of {\it tetR} ($a_i$) of all the oscillators, which
behaviors are similar to the experimental results (see, e.g.
\cite{1}). Fig. 1 (c) shows the synchronization error of $a_i-a_1$
for $i=2,\cdots,10$.

\begin{figure}[htb]
\centering
\includegraphics[width=14cm]{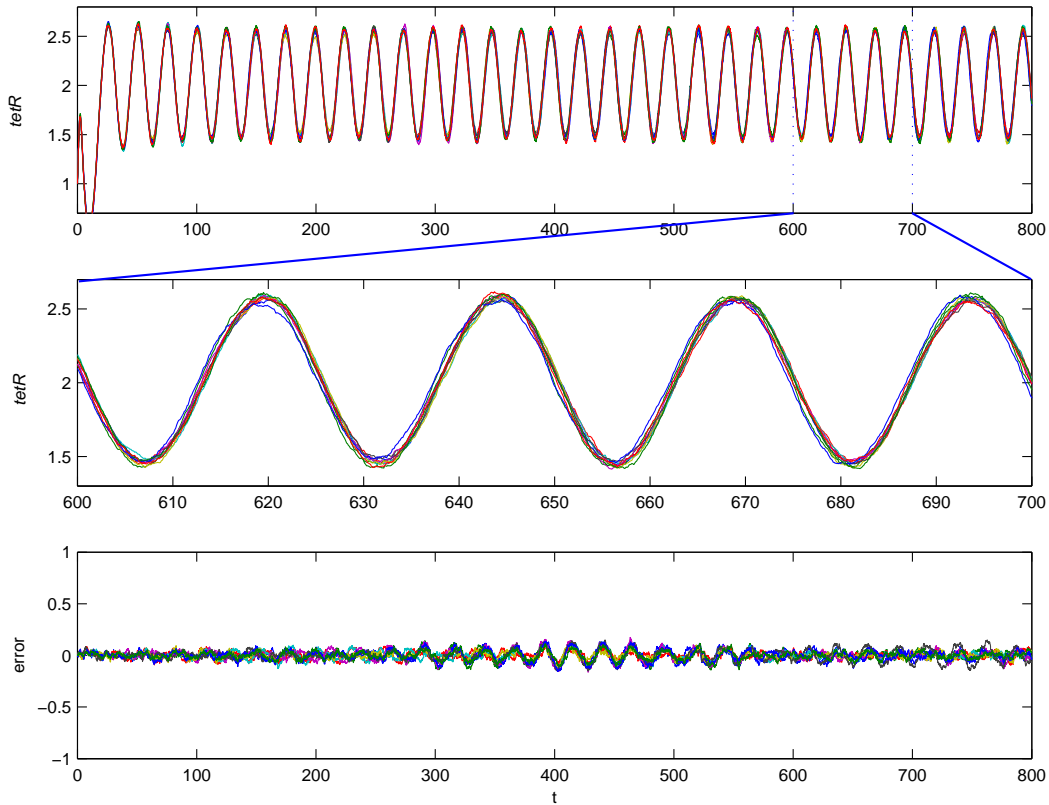}
\caption{Simulation results of the coupled repressilators with the
same initial values. (a) The evolution dynamics of the mRNA
concentrations of {\it tetR} ($a_i$) of all the genetic
oscillators. (b) Zooming in the range $t\in [600, 700]$ of (a).
(c) The evolution of the synchronization error of $a_i-a_1$ for
$i=2,\cdots,10$.}
\end{figure}
\begin{figure}[h]
\centering
\includegraphics[width=14cm]{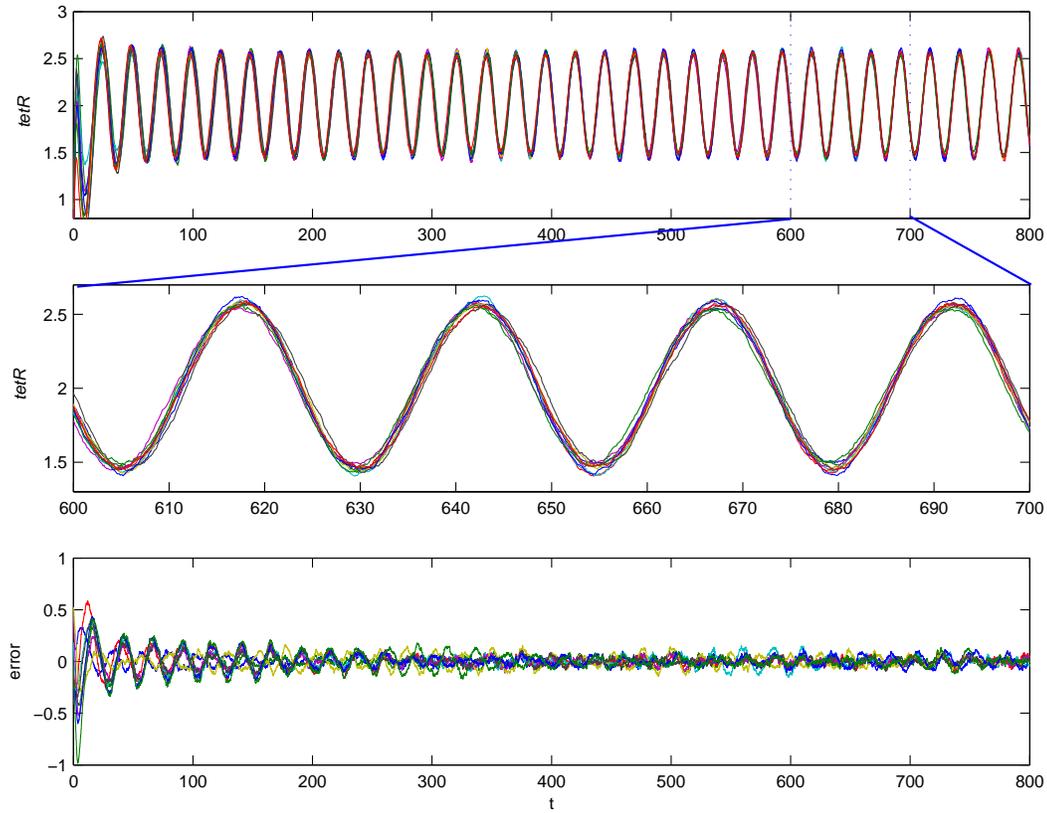}
\caption{Coupled genetic oscillators with different initial
conditions: (a) The evolution dynamics of the mRNA concentrations
of {\it tetR} ($a_i$) of all the genetic oscillators. (b) Zooming
in the range $t\in [600, 700]$ of (a). (c) The evolution of the
synchronization error of $a_i-a_1$ for $i=2,\cdots,10$.}
\end{figure}
\begin{figure}[h]
\centering
\includegraphics[width=14cm]{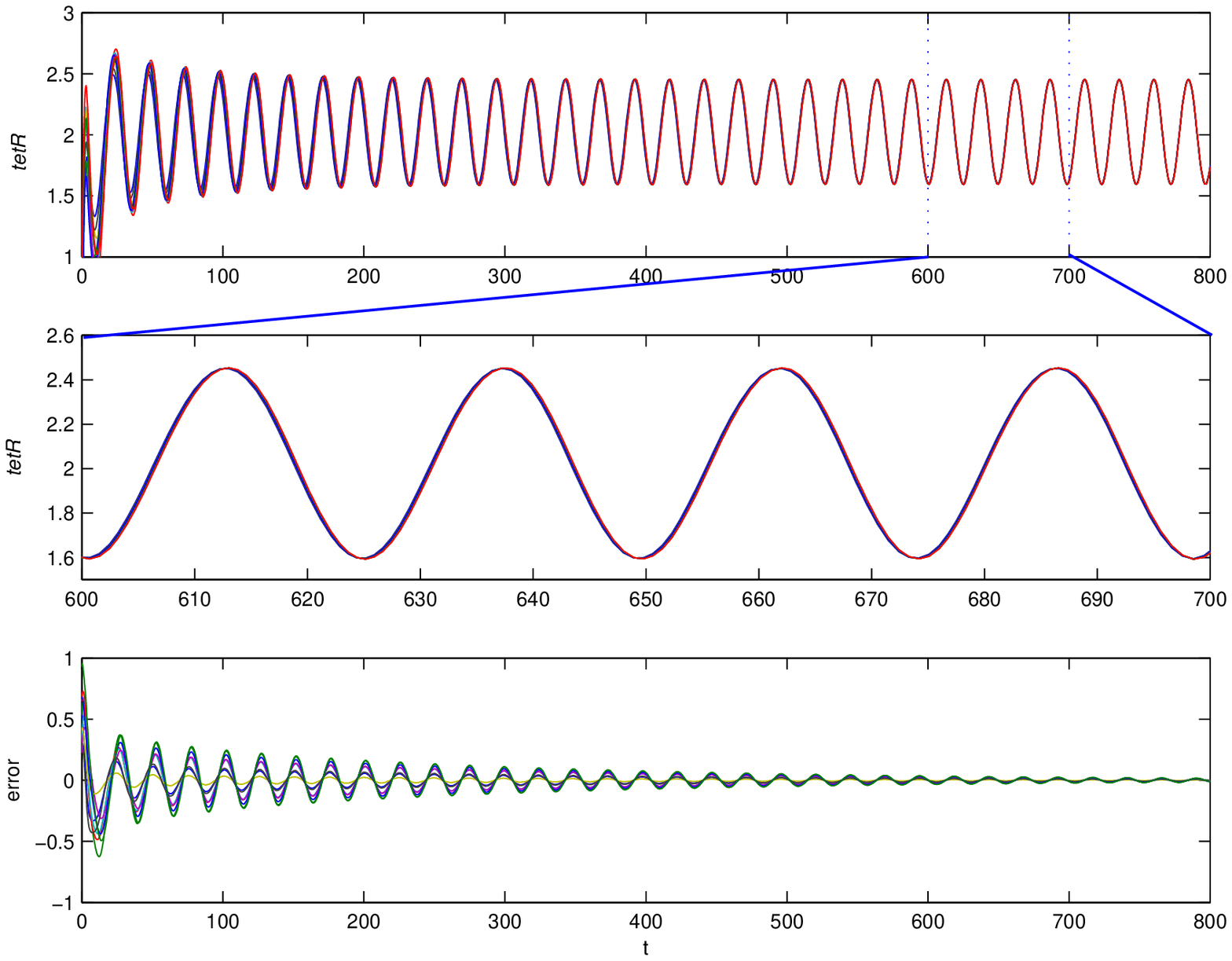}
\caption{Coupled genetic oscillators without noise perturbation:
(a) The evolution dynamics of the mRNA concentrations of {\it
tetR} ($a_i$) of all the genetic oscillators. (b) Zooming in the
range $t\in [600, 700]$ of (a). (c) The evolution of the
synchronization error of $a_i-a_1$ for $i=2,\cdots,10$.}
\end{figure}

In Definition 1, it requires that all the genetic oscillators have
the same initial conditions, so that $V(x(0))=0$. If the genetic
oscillators have different initial conditions, $V(x(0))\neq 0$,
and thus (12) in the Appendix C is replaced by
\begin{equation*}
\begin{array}{rl}
J(t)\leq & \textbf{E}\{\int_0^t[\alpha
LV(x(s))+\sum_{i<j}(-U_{ij})(x_i(s)-x_j(s))^T (x_i(s)-x_j(s))
-\gamma
\sum_iv_i^T(s)v_i(s)]ds\}\\
&+ \textbf{E}(V(x(0)).
\end{array}
\end{equation*}

Since in genetic networks, the variables usually represent the
concentrations of mRNAs, proteins and chemical complexes, which
are of (not so large) limited values, and so is $V(x(0))$. For a
long time scale, the last term of the above inequality is usually
much smaller than the absolute value of the first term in the
right-hand side, and thus the last term can be ignored roughly. In
Fig. 3, we show the same computations as those in Fig. 2 except
that the oscillators are with different initial values (randomly
in the range (0, 1)). After a period of evolution, the network
behaviors are similar to those in Fig. 2, which verifies our above
argument. In other words, rigorously, according to Definition 1,
we need that all the oscillators have the same initial conditions,
but practically, for oscillators with different initial
conditions, we can obtain almost the same results.

For the purpose of comparison, in Fig. 4 we show the simulation
results of a network without noise perturbations. As we can
conclude from Figs. 2-4, the networks with noise perturbation,
though can't achieve perfect synchronization, can indeed achieve
synchronization with small error fluctuation, and the network
behaviors are similar to those of networks without noise
perturbations.

\subsection*{Synthesis}
In addition to providing a sufficient condition for the stochastic
synchronization, Proposition 1 can also be used for designing
genetic oscillator networks, which is a byproduct of the main
results. From the synthetic biology viewpoint, to minimize the
influence of the noises (on the synchronization), we can design
genetic oscillator networks according to the following rule:
\begin{equation}
\mbox{min } \gamma, \mbox{ such that the LMIs (8) hold},
\end{equation}
which is obviously from the above theoretical result. This is
similar to the $H_\infty$ synthesis problem in control theory.

\section*{Conclusion and Outlook}
In this paper, we presented a general theoretical method for
analyzing the stochastic synchronization of coupled genetic
oscillators based on systems biology approach. By taking the
specific structure of genetic systems into account, a sufficient
condition for the stochastic synchronization was derived based on
LMI formalism, which can be easily verified numerically. Although
the method and results are presented for genetic oscillator
networks, it is also applicable to other dynamical systems. In
coupled genetic oscillator networks, since there is a maximal
activity of fully active promoters, it is more realistic to
consider a Michaelis-Menten form of the coupling terms. As argued
in \cite{5}, our theoretical method is also applicable to this
case. To make the theoretical method more understandable and to
avoid unnecessarily complicated notation, we discussed only on
some simplified forms of the genetic oscillators, but more general
cases regarding this topic can be studied in a similar way. For
example: (I) The genetic oscillator model (5) can be generalized
to more general case such that $f_{i}, g_{i}$, the component of
$f(y(t)), g(y(t))$, are functions of $y(t)$, not only of $y_i(t)$,
and $f$ and $g$ can also be of non-Hill form, provided that $0\leq
\frac{f_i(a)-f_i(b)}{c_{1i}^T(a-b)} \leq k_{1i}, 0\leq
\frac{g_i(a)-g_i(b)}{c_{2i}^T(a-b)} \leq k_{2i}, \forall a, b \in
R^n\, (a\neq b), i=1,\cdots,n$, where $c_{1i},c_{2i}\in R^n$ are
real vectors. (II) Biologically, the genetic oscillators are
usually nonidentical. We can consider genetic networks with both
parametric mismatches and stochastic perturbations in similar ways
as those presented in this paper and \cite{5}. (III) There are
significant time delays in the gene regulation, due to the slow
processes of transcription, translation and translocation. Our
result can be easily extended to the case that there are delays
both in the coupling and the individual genetic oscillators.

As we know, noises can play both beneficial and harmful roles (for
synchronization) in biological systems. For the latter case, the
noise is a kind of perturbation, and it is interesting to study
the robustness of the synchronization with respect to noise. In
this paper, we addressed this question. For the former case, in
\cite{7, plos}, the authors studied the mechanisms of
noise-induced synchronization.

\section*{Method}
To simulate the stocahstic differential equaiton
$\dot{x}(t)=f(x)+g(x)\xi(t)$, the well-known Euler-Maruyama scheme
is most frequently used, which is also used in this paper. In this
scheme, the numerical trajectory is generated by
$x_{n+1}=x_n+hf(x_n)+\sqrt{h}g(x_n)\eta_n$, where $h$ is the time
step and $\eta_n$ is a discrete time Gaussian white noise with
$<\eta_n>=0$ and $<\eta_n\eta_m>=\delta_{nm}$. For more details,
see e.g. \cite{15}.\\[1em]

\section*{Author contributions}
CL gave the topic, developed
theoretical results, designed the numerical experiments, analyzed
the data, contributed materials/ analysis tools. CL, LC and KA
wrote the paper.

\section*{Acknowledgements}
This research was supported by Grant-in-Aid for Scientific
Research on Priority Areas 17022012 from MEXT of Japan, the Fok
Ying Tung Education Foundation under Grant 101064, the Program for
New Century Excellent Talents in University, the Distinguished
Youth Foundation of Sichuan Province, and the National Natural
Science Foundation of China (NSFC) under grant 60502009.

\newpage
\section*{Appendices}
{\it A. Notations}:

Throughout this paper, $A^T$ denotes the transpose of a square
matrix $A$. The notation $M > (<)\,0$ is used to define a real
symmetric positive definite (negative definite) matrix. $R^m$
denotes the $m$-dimensional Euclidean space; and $R^{n\times m}$
denotes the set of all $n\times m$ real matrices. In this paper,
if not explicitly stated, matrices are assumed to have compatible
dimensions. $\textbf{E}(\cdot)$ denotes the expectation operator;
$L_2[0,\infty)$ is the space of square-integrable vector functions
over $[0,\infty)$; $|\cdot|$ stands for the Euclidean vector norm,
and $\|\cdot\|_2$ stands for the usual $L_2[0, \infty)$ norm. The
Kronecker product $ A \otimes B$ of an $n \times m$ matrix $A$ and
a $p \times q$ matrix $B$ is the $np \times mq$ matrix defined as
\begin{equation*}
A\otimes B=\left[\begin{array}{ccc} A_{11}B& \cdots &A_{1m}B\\
\vdots &\ddots &\vdots\\ A_{n1}B &\cdots &
A_{nm}B\end{array}\right].
\end{equation*}
For a general stochastic systems
\begin{equation*}
dx=f(t,x(t))dt+g(x(t))dw(t),
\end{equation*}
the diffusion operator $L$ acting on $Y(t,x(t))$ is defined by
\begin{equation*}
LY(t,x(t))=Y_t(t,x(t))+Y_x(t,x(t))f(t,x(t))
+\frac{1}{2}\mbox{trace}[g(x(t))g^T(x(t))Y_{xx}(t,x(t))].
\end{equation*}

{\it B. Mean-square synchronization}

For network (1), a natural attempt is to study the mean-square
asymptotic synchronization. Analogue to the definition of
mean-square stability \cite{Arnold}, we can define the mean-square
synchronization as follows:

\emph{Definition A1}: The network (1) is said to be mean-square
synchronous if for every $\epsilon >0$, there is a
$\delta(\epsilon)>0$, such that $\textbf{E}|x_i(t)-x_j(t)|_{t>0}^2
< \epsilon $ for $|x_i(0)-x_j(0)|<\delta (\epsilon), \forall i,j$.
If in addition, $\mbox{lim}_{t\rightarrow \infty}
\textbf{E}|x_i(t)-x_j(t)|^2=0$ for all initial conditions, the
network is said to be mean square asymptotically synchronous.

In analyzing the synchronization of the network (1), we use the
Lyapunov function $V(x(t))=x^T(t)(U\otimes P)x(t)$ \cite{Wu},
where $\otimes$ is the Kronecker product, and
$x(t)=[x_1^T(t),\cdots,x_n^T(t)]^T\in R^{Nn\times 1}$. According
to \cite{Wu}, this Lyapunov function is equivalent to
$V(x(t))=\sum_{i<j}(-U_{ij})(x_i(t)-x_j(t))^T P(x_i(t)-x_j(t))$.

By It$\hat{o}$'s formula \cite{Arnold}, we obtain the following
stochastic differential along (1)
\begin{equation*}
dV(x(t))=LV(x(t))dt+2x^T(t)(U\otimes P)v(t)dw(t)
\end{equation*}
where $v(t)=\mbox{diag}(v_1,\cdots,v_N)\in R^{Nn\times N}$, $L$ is
the diffusion operator, and
\begin{equation*}
\begin{array}{rl}
LV(x(t))=&2\sum_{i<j}(-U_{ij})(x_i-x_j)^T
P[F(x_i)-F(x_j)-T(x_i-x_j)]\\
&+2x^T(t)(U\otimes P)(G\otimes D+I\otimes
T)x(t)\\&+\mbox{trace}(v(t)v^T(t)(U\otimes P))
\end{array}
\end{equation*}

We discuss two special cases of the stochastic terms:

1. The genetic oscillators are perturbed by the same noise, which
can occur in the situation that genetic oscillators communicate
via a common environment. In this case, $v_i dw_i$ are the same
for all $i$. We let $v=[v_1^T,\cdots,v_N^T]^T$ and $dw=dw_i$.
Since $U$ is a matrix with zero row sums and $v_i$ is the same for
all $i$, it is easy to show that the last term of $LV$ is zero.
Thus if the following conditions hold, we will have
$\textbf{E}[dV(x(t))] =\textbf{E}[LV(x(t))dt]<0$.
\begin{equation*}
\begin{array}{c}
(y_1-y_2)^TP[F(y_1)-F(y_2)-T(y_1-y_2)]<0,\\ \forall y_1,y_2 \in R^n\, (y_1\neq y_2)\\
(U\otimes P)(G\otimes D+I\otimes T)+(G\otimes D+I\otimes
T)^T(U\otimes P) \leq 0.
\end{array}
\end{equation*}
Hence, if there are matrices $P>0, T\in R^{n\times n}$ and $U$,
such that the above conditions hold, the network (1) will achieve
mean-square asymptotically synchronization. In this case, roughly
speaking, the noise will not affect the synchronous state (since
they are common for all oscillators), but it will affect the
individual oscillator dynamics.

2. The noise intensity matrix $v_i$ is a function of
$\sum_jG_{ij}Dx_j$, which means that if there is no coupling from
oscillator $j$ to $i$, then $j$ does not have contribution to the
perturbation of oscillator $i$. We further assume that $v_i$ can
be estimated by
\begin{equation*}
v_i^Tv_i \leq
\left[\sum_jG_{ij}Dx_j(t)\right]^TH_{i}\left[\sum_jG_{ij}Dx_j(t)\right],\hspace{0.1cm}
H_{i}\geq 0.
\end{equation*}
Defining $v=\mbox{diag}(v_1,\cdots,v_N)\in R^{Nn\times N}$,
$w(t)=[w_1(t),\cdots,w_N(t)]^T$, and
$H=\mbox{diag}(H_1,\cdots,H_N)$, and assuming $U\otimes P\leq \rho
I$, we have
\begin{equation*}
\begin{array}{rl}
\mbox{trace}(v(t)v^T(t)(U\otimes P)) &\leq\lambda_{max}(U\otimes
P) \mbox{trace}(v(t)v^T(t))\\&\leq\rho\,
\sum_iv_i^T(t)v_i(t)\\
&\leq \rho x^T(t)(G\otimes D)^TH(G\otimes D)x(t).
\end{array}
\end{equation*}
So, the conditions for the mean-square asymptotically
synchronization of the network (1) in this case are
\begin{equation*}
\begin{array}{c}
(y_1-y_2)^TP[F(y_1)-F(y_2)-T(y_1-y_2)]<0,\\ \forall y_1,y_2 \in R^n\, (y_1\neq y_2)\\
(U\otimes P)(G\otimes D+I\otimes T)+(G\otimes D+I\otimes
T)^T(U\otimes P)+\rho(G\otimes D)^TH(G\otimes D)\leq 0,\\
U\otimes P\leq \rho I.
\end{array}
\end{equation*}

If we consider genetic oscillators of the form of (5), the
conditions for the mean-square asymptotically synchronization can
be analyzed by the same method as that in the following Appendix
D.

From Definition A1, we know that the definition of the mean-square
asymptotically synchronization is rather restrictive, which
requires that $\mbox{lim}_{t\rightarrow \infty} \textbf{E}
|x_i(t)-x_j(t)|^2=0, \forall i, j$. If it is neither of the above
two cases, that is neither $v_i dw_i$ are the same for all $i$,
nor $v_i\,(\forall i)$ reduce to zero when $x_1=\cdots=x_n$, the
network is hardly to achieve mean-square asymptotically
synchronization. Experimental results also show that usually the
genetic oscillators can not achieve mean-square synchronization
(see for example \cite{1}). So, we argue that the study of
mean-square synchronization is unrealistic (and therefore
meaningless) in genetic networks.

In Ref. \cite{Lin}, the authors studied the mean-square asymptotic
synchronization of two master-slave coupled Chua's circuits. They
assume that the noise intensity depends on the difference of the
states of the two systems, which is also somewhat unrealistic.

\emph{C. Analysis of the general synchronization condition}

To obtain the general synchronization condition (3) of the network
(1), we also use the Lyapunov function $V(x(t))=x^T(t)(U\otimes
P)x(t)$. By It$\hat{o}$'s formula \cite{Arnold}, we obtain the
stochastic differential $dV(x(t))=LV(x(t))dt+2x^T(t)(U\otimes
P)v(t)dw(t)$. According to Definition 1, we assume that the
oscillators have the same initial conditions, thus we can derive $
\textbf{E}(V(x(t))=\textbf{E}\left(\int_0^tLV(x(s))ds\right)$. For
$\gamma>0$, we define
\begin{equation}
J(t)=\textbf{E}\{\int_0^t[\sum_{i<j}(-U_{ij})(x_i(s)-x_j(s))^T(x_i(s)-x_j(s))
-\gamma \sum_iv_i^T(s)v_i(s)]ds\}
\end{equation}
Then, it is easy to show that for $\alpha>0$,
\begin{equation}
\begin{array}{rl}
J(t)\leq & \textbf{E}\{\int_0^t[\alpha
LV(x(s))+\sum_{i<j}(-U_{ij})(x_i(s)-x_j(s))^T\\&
\cdot(x_i(s)-x_j(s)) -\gamma
\sum_iv_i^T(s)v_i(s)]ds\}\\
\equiv &\textbf{E}\{\int_0^tS_1(s)ds\}.
\end{array}
\end{equation}

Assuming $U\otimes P\leq \rho I$, and letting
$\alpha=\gamma/\rho$, we have
\begin{equation*}
\begin{array}{rl}
S_1=&\frac{\gamma}{\rho}\{2\sum_{i<j}(-U_{ij})(x_i(t)-x_j(t))^T\\
&\cdot P[F(x_i(t))-F(x_j(t))-T(x_i(t)-x_j(t))]\\
&+2x^T(t)(U\otimes P)(G\otimes D+I\otimes T)x(t)\\
&+\mbox{trace}(v(t)v^T(t)(U\otimes P))\\
&+\frac{\rho}{\gamma}\sum_{i<j}(-U_{ij})(x_i(t)-x_j(t))^T
(x_i(t)-x_j(t))\\
&-\rho\sum_iv_i^T(t)v_i(t)\}\\
\leq & \frac{\gamma}{\rho}\{\sum_{i<j}(-U_{ij})[2(x_i(t)-x_j(t))^T\\
&\cdot P(F(x_i)-F(x_j)-T(x_i-x_j))\\
&+\frac{\rho}{\gamma}(x_i(t)-x_j(t))^T (x_i(t)-x_j(t))]\\
&+2x^T(t)(U\otimes P)(G\otimes D+I\otimes T)x(t)\}.
\end{array}
\end{equation*}
If $\textbf{E}(S_1)<0$, we will have $J(t)<0$, and thus, (2)
follows immediately from (3).

\emph{D. Proof of Proposition 1}

Proposition 1 can be proved by replacing $F$ in $S_2$ of (3) by
the dynamics of (5), and using the sector conditions (6). We have
\begin{equation*}
\begin{array}{rl}
S_2=&2(y_1(t)-y_2(t))^T
P[(A-T)(y_1(t)-y_2(t))\\&+B_1(f(y_1(t))-f(y_2(t)))-B_2(g(y_1(t))-g(y_2(t)))]\\
&+\frac{\rho}{\gamma}(y_1(t)-y_2(t))^T(y_1(t)-y_2(t))\\
\leq &2(y_1(t)-y_2(t))^T
P(A-T)(y_1(t)-y_2(t))\\& +\frac{\rho}{\gamma}(y_1(t)-y_2(t))^T(y_1(t)-y_2(t))\\
&+2(y_1(t)-y_2(t))^TPB_1(f(y_1(t))-f(y_2(t)))\\
& -2(y_1(t)-y_2(t))^TPB_2(g(y_1(t))-g(y_2(t)))\\
&-2\sum_{l=1}^n\lambda_{1l}(f_l(y_{1l}(t))-f_l(y_{2l}(t)))\\
&\cdot [(f_l(y_{1l}(t))-f_l(y_{2l}(t)))-k_1(y_{1l}(t)-y_{2l}(t))]\\
&-2\sum_{l=1}^n\lambda_{2l}(g_l(y_{1l}(t))-g_l(y_{2l}(t)))\\&\cdot
[(g_l(y_{1l}(t))-g_l(y_{2l}(t)))-k_2(y_{1l}(t)-y_{2l}(t))].
\end{array}
\end{equation*}
By letting $Q=PT$ and denoting the first matrix in (8) by $M_1$,
we have $S_2\leq\xi(t)M_1\xi(t)<0$ for all $y_1,y_2$ except for
$y_1=y_2$, where
$\xi(t)=[(y_1(t)-y_2(t))^T,(f(y_1(t))-f(y_2(t)))^T,
(g(y_1(t))-g(y_2(t)))^T]^T\in R^{3n\times 1}$. So, the first
condition in (3) is satisfied. Substituting $Q=PT$, the second
inequality in (3) is equivalent to the second inequality in (8).
Thus, Proposition 1 is proved.

\end{document}